\documentclass[12pt]{article}

\newcommand{\bea}{\begin{eqnarray}}
\newcommand{\ba}{\begin{array}}
\newcommand{\ea}{\end{array}}
\newcommand{\eea}{\end{eqnarray}}

\newcommand{\be}{\begin{equation}}
\newcommand{\ee}{\end{equation}}

\newcommand{\scs}{\scriptstyle}
\begin{document}

\begin{center}

{\large \textbf{\ A Covariant Generalization of the Real-Time Green's
Functions Method \\[0pt]
in the Theory of Kinetic Equations\\[0pt]
}} \vspace*{1.2cm} {S.~A.~Smolyansky and A.~V.~Prozorkevich} \\[0pt]
\emph{Physics Department, Saratov State University, 410071, Saratov, Russia}%
\\[0pt]

\vspace*{.3cm}

{G.~Maino}\\[0pt]
\emph{ENEA, Applied Physics Division, via Don G. Fiammelli 2, 40129,
Bologna, Italy}\\[0pt]

\vspace*{.3cm}

{and}\\[0pt]

\vspace*{.3cm}

{S.~G.~Mashnik }\\[0pt]
\emph{T-2, Theoretical Division, Los Alamos National Laboratory, Los Alamos,
NM 87545}\\[0pt]

\end{center}

\vspace*{10pt}

\abstract{\ }

A generalized quantum kinetic equation (RKE) of the Kadanoff-Baym type is
obtained on the basis of the Heisenberg equations of motion where the time
evolution and space translation are separated from each other by means of
the covariant method. The same approach is used also for a covariant
modification of the real-time Green's functions method based on the Wigner
representation. The suggested approach does not contain arbitrariness'
elements and uncertainties which often arise from derivation of RKE on the
basis of the motion equations of the Kadanoff-Baym type for the correlation
functions in the case of systems with inner degrees of freedom.

Possibilities of the proposed method are demonstrated by examples of
derivation of RKE of the Vlasov type and collision integrals of the
Boltzmann-Uehling-Uhlenbeck (BUU) type in the frame of the $\sigma \omega$%
-version of quantum hadrodynamics, for the simplest case of spin saturated
nuclear matter without antinuclear component. Here, the quasiparticle
approximation in a covariant performance is used. A generalization of the
method for the description of strong non-equilibrium states based on the
non-equilibrium statistical operator is then proposed as well.

\section{Introduction}

The recent growing interest in the relativistic nuclear physics, especially
at intermediate energies, has stimulated the development of the theory of
relativistic kinetic equations (RKE) as well. In particular, the main
problems of dynamical RKE derivation were discussed within the framework of
different approaches with well grounded traditions in the non-relativistic
case, namely the method of the real-time (contour) Green's functions \cite
{BM}-\cite{VB}, the BBGKY method (method of many particles correlation
functions) \cite{CM,W}, the method of the non-equilibrium statistical
operator \cite{S1,S2}, and several other techniques \cite{CH}-\cite{DM}.

However, in some approaches, the transition to the relativistic region it is
not obvious and can lead to misleading results. This situation was analyzed
in Refs.\cite{S3}-\cite{S5} on a rather simple quantum field model which
describes interaction of the fermion and boson subsystems in the mean field
approximation. In these works, it was shown that the equations of motion for
a two-point correlation function lead to true RKE of the Vlasov type only if
one uses some definite simple rules of projection. It is important that the
same RKE can be derived also by a direct method based on the Heisenberg
equation of motion \cite{S1,S2}, \cite{S3}-\cite{S10}.
 In the latter approach, any ambiguities in the RKE derivation
are thus in general avoided. These features of the direct
method represent a suitable expedient to obtain  a correct generalized RKE in
the frame of a relativistic analogue of the Kadanoff-Baym
formalism. This is the main result of our study.

Above--mentioned works \cite{S1,S2}, \cite {S3}-\cite{S10}
are devoted to covariant generalizations of the Zubarev method
of non-equilibrium
statistical operator \cite{Z-book} in terms of the relativistic Wigner
functions.  However, this formalism presents some technical
difficulties to go beyond the Born approximation, while the
formalism introduced in this work, based on the real--time Green function,
avoids this kind of mathematical troubles.

In order to solve this main problem, we develop a specific relativistic
modification of the real-time Green's functions method in the Wigner
representation. This improvement proves to be possible since the transition
to the Wigner representation in relativistic theory allows us to introduce a
time-like direction which can be connected with the momentum argument of the
Wigner function. As a result, the time variable determining the evolution of
the system can be defined as a true scalar, which plays the role of the
proper time in a given point of the eight-dimensional phase space. This
invariant time is used also to define the chronological ordering operation
in the real-time Green's functions . It leads to an obvious modification of
this method at all stages of the formalism including the description of the
dispersion properties of quantum field systems. Such a modification of the
real-time Green's functions method is convenient also in order to describe
the dispersion properties of a quantum-field system at finite temperature
and density as it proves to be obviously relativistic invariant for all the
steps of the relevant calculations.

The outline of the paper is as follows. In Sec. 2, we discuss shortly the
''direct way'' of the derivation of RKE in terms of one-particle
relativistic Wigner's functions. The advantages of this approach are its
simplicity and clearness. This fact originates from the use of the
Heisenberg picture from the very beginning. However, in this section, we do
not discuss any truncation scheme and derivation of closed-form RKE. This
choice is convenient for a comparison with the method based on the
Kadanoff-Baym motion equations for correlation functions (Sec. 3). A
covariant modification of the real-time Green's functions method is found as
an adequate formalism for this aim. As a result, we obtain a generalized RKE
of the Kadanoff-Baym type without using any irrelevant assumption, thus
allowing us to use effectively the suggested method for a reliable kinetic
description of a quite wide class of systems with inner degrees of freedom.

In essence, the only limitation of the suggested method is a request of a
polynomial character of interaction of systems under study. It is important
that the suggested approach results in a self-consistent, non-contradictory,
and unambiguous theory of kinetic equations. To illustrate the possibilities
of the theory, two examples are discussed in detail. Firstly, a RKE of the
Vlasov type (the mean field approximation, Sec. 2) and then a collision
integral of the BUU type (the quasiparticle approximation, Sec. 4). The
standard Walecka model, which is a quantum field model of the relativistic
nuclear matter consisting of nucleons and two types (scalar and vector) of
mesons (quantum hadrodynamics) is used in our work as a toy model.

In Sec. 5, we consider another generalization of relativistic kinetic theory
based on a change of the standard averaging procedure under the equilibrium
density matrix by an averaging using the non-equilibrium statistical
operator. This step leads to a possible kinetic description of strongly
non-equilibrium states which are often met in the relativistic nuclear
physics. Finally, conclusions are drawn in Sec. 6. Everywhere, we work with
natural units, $\hbar =c=1$c.

\section{ The motion equation for the Wigner function}

To illustrate our approach on a nontrivial example of a system with internal
degrees of freedom, let us consider the kinetic description of the Fermi
subsystem of a quantum-field system. For a consistent dynamical construction
of the relativistic kinetic theory, we will start with the introduction of
the one-particle Wigner function of the Fermi subsystem,
\begin{equation}  \label{10}
f_{\alpha \beta }(x,p)=(2\pi )^{-4}\int dy\, e^{-ipy}<P_{\alpha \beta
}(x,y)> \mbox{ ,}
\end{equation}
where the symbol $<...> = \mathrm{Tr}...\rho $ denotes the operation of
statistical averaging with density matrix $\rho $ in the Heisenberg
representation and
\begin{equation}  \label{20}
P_{\alpha \beta }(x,y)=\bar \psi _\beta (x_+)\psi _\alpha (x_-) \mbox{,}
\end{equation}
$\psi(x)$ and $\bar \psi (x)$ being the usual field operators, and $x_\pm
=x\pm y/2$.

In the realm of dynamical basis of the theory, we choose the Heisenberg
equations of motion, namely,
\begin{equation}  \label{30}
i\partial ^\mu A(x)=[A(x),P^\mu ] \mbox{ ,}
\end{equation}
where $A(x)=A[x,\psi(x),\bar\psi(x)]$ is an arbitrary local operator and $%
P^\mu$ is the total 4-momentum of the system,
\begin{equation}  \label{40}
P^\mu = \int d\sigma _\nu (x\!\!\mid\!\!n)T^{\mu \nu} \mbox{ .}
\end{equation}
Here, $d\sigma _\nu (x\!\mid\! n)$ is a vector element of an arbitrary
space-like hyperplane, $\sigma (n),$ with a time-like normal vector, $n^\mu
(n^2 =1),$ and $T^{\mu \nu}= T^{\nu \mu}$ is the energy-momentum tensor..

The energy and momentum conservation laws lead to the independence of the
integrals (\ref{40}) from the selection of a given hyperplane  $\sigma (n)$ ,
and  we
can fix the time-like direction by means of an external condition (a
relevant example will be given below).

In order to describe the dynamics of the system along the time-like
direction, $n^\mu,$ and the space translation along the independent
space-like directions on the hyperplane, $\sigma(n)$, a boost transformation
with the $^{\prime\prime}$velocity$^{\prime\prime}$, $n^\mu, $ can be used
for the motion Eqs. (\ref{30}). Technically, this result is achieved using
the projection of Eqs. (\ref{30}) on the direction $n^\mu$ and on the
hyperplane, $\sigma(n)$. Convolution of Eqs. (\ref{30}) with the vector, $%
n^\mu ,$ leads to the following dynamical equation \cite{n},
\begin{equation}  \label{50}
i\, \frac {\partial A(x)}{\partial \tau }=[A(x),H(\tau )] \mbox{ .}
\end{equation}
The parameter,
\begin{equation}  \label{60}
\tau =n_\mu x^\mu ,
\end{equation}
plays here the role of a proper time in a new coordinate system. In Eq.(\ref
{50}), the derivative along the direction of $n^\mu$ is thus introduced,
\begin{equation}  \label{70}
\frac {\partial }{\partial \tau }=n^\mu \frac {\partial }{\partial x^\mu }
\quad \mbox{.}
\end{equation}
$H(\tau )$ is the scalar Hamiltonian of the system and reads as
\begin{equation}  \label{80}
H(\tau )=n_\mu P^\mu=\int d\sigma (n)n_\mu T^{\mu \nu }n_\nu \mbox{,}
\end{equation}
because $d\sigma ^\mu (x\!\mid\! n)=n^\mu d\sigma (x\!\mid\! n) $. This
procedure of a formal restoration of the relativistic invariance in the
description of the time evolution is very popular in the quantum field
theory (see, e.g., \cite{n}).

The projection of Eqs. (\ref{30}) on the space-like directions
is carried out by the convolution of Eqs. (\ref{30}) with the
projection operator,
\begin{equation}  \label{90}
\Delta ^{\mu \nu }=g^{\mu \nu }-n^\mu n^\nu ,\quad n_\mu
\Delta ^{\mu \nu }=0 \mbox{ .}
\end{equation}
The resulting equations provide a description of infinitesimal
transforms of the system on the space-like hyperplane, $\sigma
(n)$. These equations are
not interesting for the present aim of developing a kinetic theory \cite{S3}-%
\cite{S5}.

Now, we can write the motion equation for the Wigner function
(\ref{10}). Performing the $\tau$-differentiation of function
(\ref{10}) and using the Liouville equation in the Heisenberg
representation,
\begin{equation}  \label{100}
\frac{d\rho}{d\tau}=0 \mbox{ ,}
\end{equation}
after substitution of the motion equation (\ref{50}) and using
the definition (\ref{70}), we find:
\begin{equation}  \label{110}
\frac {\partial f(x,p)}{\partial \tau }=n^\mu
\frac{\partial}{\partial x^\mu} f(x,p)=-i(2\pi )^{-4}\int dy
e^{-ipy}<[P(x,y),H(\tau )]> \mbox{ .}
\end{equation}
It is worth remarking that, for sake of convenience, the spin
indices are dropped out in our notation.

At this stage, we can remove the arbitrariness in the choice of the unit
time-like vector, $n^\mu,$ using the specific character of the mixed Wigner$%
^{\prime}$s representation. Let us assume that the momentum
vector, $p^\mu,$ in Eq. (\ref{110}) is a time-like one
(Assumption 1). Then, we can define a unit
vector in the direction of $p^\mu$ and identify it with the vector, $n^\mu$%
(Assumption 2) \cite{S1,S2}, \cite{S3}-\cite{S5}:
\begin{equation}  \label{120}
n^\mu \stackrel{def}{=}u^\mu =p^\mu /\sqrt{p^2} \mbox{ ,}
\quad u^2=1 \mbox{.}
\end{equation}

Assuming a description of non-equilibrium quantum field
systems in terms of the Wigner functions (\ref{10}), often one
limits oneself to consider only processes on the mass shell,
\begin{equation}  \label{130}
p^2=m^2 \mbox{ .}
\end{equation}
In this respect, introduction of Assumptiones 1 and 2 does not
produce any essential restriction even at the off mass shell.
In other words, Assumption 1 extracts the physical significant
domain, $\mathcal{P}^+,$ from the full phase space,
$\mathcal{P}\equiv (x,p)$. Let us remark that an analogous
restriction is usually introduced at the stage of the definition of Wigner$%
^{\prime}$s functions of the states with positive and negative
energies \cite {G}. On the other hand, Assumption 1 can be
taken off and go out into the space-like region (see the
discussion at the end of Sec. 3). We conserve this restriction
here in order to remain into the framework of the ordinary
dynamical description.

The effectiveness of the introduction of Assumptiones 1 and 2
is confirmed by the fact that substitution of the relation
(\ref{120}) into Eq. (\ref{110})
completely removes any arbitrariness in this equation and leads to a $%
^{\prime\prime}$generalized RKE$^{\prime\prime}$,
\begin{equation}  \label{140}
\begin{array}{c}
p_\mu \partial ^\mu (x) f(x,p)=-i(2\pi )^{-4}\sqrt{p^2}\int dy
e^{-ipy} <[P(x,y),H(\tau )]>= \vphantom{\biggl ]} \\ =-i(2\pi
)^{-4}\sqrt{p^2}\int dy e^{-ipy} <[\bar \psi
(x_+),H(\tau_+)]\psi (x_-) + \bar \psi (x_+)[\psi
(x_-),H(\tau_-)]> \ ,\vphantom{\biggl ]}
\end{array}
\end{equation}
where we took into account the arbitrariness of the assigned
hyperplane into the Hamiltonian (\ref{80}), $H(\tau) =
H(\tau_+) = H(\tau_-)$. Essentially, this is the first
equation of the BBGKY hierarchy in the Wigner representation.
In the mean field approximation, Eq. (\ref{140}) results to be
a non-contradictory RKE of the Vlasov type
\cite{S3}-\cite{S5}, \cite {S9,S10}.

In a general case, it is necessary to introduce different
truncation schemes to obtain closed-form RKE from Eq.
(\ref{140}). The usual perturbation theory on the coupling
constant starting from Eq. (\ref{140}) was developed in
Refs.~\cite{S1,S2}, \cite{S6}-\cite{S8}. Within the framework
of such a perturbation theory, two types of RKE with collision
integrals of the second
order have been obtained: RKE of the Bloch type (for the vertices of the $%
^{\prime\prime}$three-tails$^{\prime\prime}$ type) and RKE of
the Boltzmann
type (for the effective vertices of the $^{\prime\prime}$four-tails$%
^{\prime\prime}$ type). The Green$^{\prime}$s functions method
offers an effective way to go beyond the perturbation
approach. In the following Section, Eq.(\ref{140}) will be
re-formulated in terms of modified contour Green$^{\prime}$s
functions. In particular, this will allow us to get a
unambiguous generalized RKE of the Kadanoff-Baym type, thus
opening some perspectives to go beyond the standard
perturbation theory without breakdown of the relativistic
invariance of kinetic theory.

\section{ Generalized RKE in terms of modified \\
real-time Green$^{\prime}$s functions}

\subsection{Covariant real-time Green$^{\prime}$s functions}

The modern technique of derivation of the Kadanoff-Baym type
kinetic equations is based on the real-time Green$^{\prime}$s
functions method \cite
{BM,SC,MH,VB,Z2,D}. Now, 
our aim is to combine together the $^{\prime\prime}$
generalized RKE$^{\prime\prime}$(\ref{140}) and a system of
motion equations of the Kadanoff-Baym type for the real-time
Green$^{\prime}$s functions. As a first step, we will modify
this method taking into account the existence of a preferred
time-like direction (Assumptiones 1 and 2 in Sec. 2) in the
definition of the Green$^{\prime}$s functions in the Wigner
representation. This direction is fixed by relation
(\ref{120}) and leads to scalar time variables (see Eq.
(\ref{60})),
\begin{equation}  \label{150}
\tau _i=x^\mu _i n_\mu,\quad n^\mu \to u^\mu \mbox{ ,}
\end{equation}
where $x^\mu _i$ are arguments of the two-points
Green$^{\prime}$s
functions. 
It is implied that identification (\ref{120}) is fulfilled
after a transition to the Wigner representation. The time
arguments (\ref{150}) are convenient for use in the definition
of the ordering operation determining the modified real-time
Green$^{\prime}$s functions. Limiting ourselves to
the case of a Fermi subsystem, let us introduce the following Green$%
^{\prime} $s functions in the Wigner representation,
\begin{equation}  \label{160}
G^{(k)}_{\alpha \beta}(xp)=(2\pi)^{-4}\int
dy\,e^{ipy}\,G^{(k)}_{\alpha \beta}(x_+,x_- ) \mbox{ ,}
\end{equation}
where index (k) enumerates the elements of the real-time (contour) Green$%
^{\prime}$s function,
\begin{equation}  \label{170}
\begin{array}{lclcl}
G^c_{\alpha\beta}(x_1,x_2) & = & -i<T^c [\psi
_{\alpha}(x_1)\bar \psi _{\beta} (x_2)]> & = & G^{--}_{12}
\mbox{ ,}\vphantom{\biggl ]} \\ G^a_{\alpha\beta}(x_1,x_2) & =
& -i<T^a [\psi _{\alpha}(x_1)\bar \psi _{\beta}(x_2)]> & = &
G^{++}_{12} \mbox{ ,}\vphantom{\biggl ]} \\
G^>_{\alpha\beta}(x_1,x_2) & = & -i<\psi _{\alpha}(x_1)\bar
\psi _{\beta} (x_2)> & = & G^{+-}_{12} \mbox{
,}\vphantom{\biggl ]} \\ G^<_{\alpha\beta}(x_1,x_2) & = &
\phantom{+}i<\bar \psi _{\beta} (x_2)\psi _{\alpha}(x_1)> & =
& G^{-+}_{12} \mbox{ ,}\vphantom{\biggl ]}
\end{array}
\end{equation}
and $(T^a)T^c$ represents (anti)chronological time ordering along the $\tau$%
-axis:
\begin{equation}  \label{180}
\begin{array}{lcl}
T^c[A(x_1)B(x_2)] & \stackrel{def}{=} & \theta\vphantom{\biggl
]} (\tau _1-\tau_2)A(x_1)B(x_2)- \theta (\tau
_2-\tau_1)B(x_2)A(x_1) \mbox{ ,} \\ T^a[A(x_1)B(x_2)] &
\stackrel{def}{=} & \theta (\tau _2-\tau_1)A(x_1)B(x_2)
-\theta (\tau _1-\tau_2)B(x_2)A(x_1) \mbox{ .}\vphantom{\biggl
]}
\end{array}
\end{equation}
The operators, $A(x)$and $B(x),$ are selected from the set of
the field operators, $\bar \psi(x)$ and $\psi(x)$.

The path-ordered real-time contour Green$^{\prime}$s functions
are defined by the relation
\begin{equation}  \label{280}
G(x_1,x_2)=-i<T_c [\psi (x_1)\bar \psi (x_2)]> \mbox{ ,}
\end{equation}
where $T_c$ is the path-ordering operator on the
$\tau$-contour which is chosen in the usual way (see Fig. 1).

\begin{picture}(160,25)(0,5)
\put(10,15){\line(1,0){100}} 
\put(60,5){\line(0,1){20}}
\thicklines
\put(65,18){\vector(1,0){5}}
\put(80,12){\vector(-1,0){5}}
\put(20,18){\line(1,0){80}}
\put(20,12){\line(1,0){80}} 
\put(100,15){\oval(6,6)[r]}
\put(20,15){\oval(6,6)[l]}
\put(22,21){\makebox(0,0){$\tau_i$}}
\put(102,21){\makebox(0,0){$\tau_f$}}
\put(112,12){\makebox(0,0){$Re\,\tau$}}
\put(66,24){\makebox(0,0){$Im\, \tau$}}
\end{picture}

\begin{center}

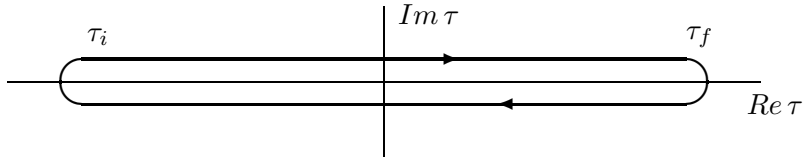
Fig. 1. The Keldysh-Schwinger contour, $C,$ with the upper $C^+$ and lower $%
C^- $ branches.
\end{center}

It is worth stressing that this definition is based on the
Wigner representation (\ref{160}) and Assumptiones 1 and 2.
This restriction leads by itself to a break-down of a simple
analogy between the Wigner and the Fourier transforms.

Let us return to Eq. (\ref{140}). From the definitions,
(\ref{10}) and (\ref {160},\ref{170}), it follows that
\begin{equation}  \label{190}
G^<(xp)=if(xp) \mbox{ .}
\end{equation}
Therefore, Eq. (\ref{140}) can be rewritten in the form
\begin{equation}  \label{200}
p^\mu\partial_\mu(x)G^<(xp)=(2\pi)^{-4}\sqrt {p^2}\int dy
e^{-ipy}<[P(x,y),H(\tau )]>\mbox{ .}
\end{equation}

Let us assume now that the Hamiltonian of the fermion
subsystem has a polynomial structure and adopt the usual
notation,
\begin{equation}  \label{210}
H(\tau)= H_0(\tau) +H_{in}(\tau),\quad H_{in}(\tau)=H_{MF}(\tau)+ H_r(\tau) %
\mbox{ .}
\end{equation}
Here, $H_0(\tau)$ describes the free evolution of the fermion
field,
\begin{equation}  \label{220}
H_0(\tau)=-\int d\sigma(x\!\mid\! u)\bar \psi (x)\{\frac i2
\gamma^\mu \stackrel{\leftrightarrow}{\partial ^\perp _\mu
}(x)-m\}\psi(x)\mbox{ ,}
\end{equation}
where $\partial ^\perp _\mu(x) $ is the space-like derivative,
\begin{equation}  \label{230}
\partial ^\perp _\mu (x)=\Delta^\nu_\mu \partial _\nu (x)\mbox{ ,}
\end{equation}
and $\Delta^{\mu\nu}$ is the projection operator (\ref{90}).
In the interaction Hamiltonian, $H_{in}(\tau),$ Eq.
(\ref{210}), the $H_{MF}(\tau)$ part, corresponding to the
mean field approximation, is explicitly separated. Then,
$H_r(\tau)$ describes the residual part of the interaction.

It is easy to calculate the free motion part in the right-hand side of Eq. (%
\ref{200}) for the Hamiltonian (\ref{220}). For covariant
transformations, it is convenient to use the following rule,
\begin{equation}  \label{240}
\int d\sigma (x^{\prime}\!\mid\!
u)S(x-x^{\prime})\,y(x^{\prime})=iu_\mu \gamma ^\mu y(x),\quad
x\in \sigma (u) \mbox{ ,}
\end{equation}
where $y(x)$ is an arbitrary function of field operators and $
S(x)$ is the anti-commutation function of the fermion fields,
namely $ S(x-x')=i[\psi (x),\bar\psi (x')]_+\ .$
 Relation
(\ref{240}) is based on a covariant generalization of the
well-known property of the function $S(x)$  (see, e.g.,
\cite{G})
\begin{equation}  \label{250} S(x)\mid _{x^0=0}=i\gamma ^0
\delta ^{(3)}(x) \mbox{ ,} \end{equation} (a simple proof of
Eq. (\ref{240}) is given in Appendix A). This kind of
calculation for Eq. (\ref{200}) leads to the following result:
 \begin{equation}  \label{260}
\begin{array}{l} p^\mu\partial_\mu (x)G^< (xp)+\frac 12 [\hat p\gamma^\mu,
\partial_\mu (x)G^< (xp)]+im[\hat p,G^< (xp)]=\vphantom{\biggl ]} \\
=(2\pi)^{-4}\sqrt{p^2}\int dy e^{-ipy}
<[\bar\psi(x_+),H_{in}(\tau_+)]\psi(x_-)+\bar\psi(x_+)[\psi(x_-),
H_{in}(\tau_-)]> \mbox{ ,}\vphantom{\biggl ]}
\end{array}
\end{equation}
where $\hat p =p^\mu \gamma_\mu $.

The following step consists in the search of a connection
between the right-hand side of Eq. (\ref{260}) and the mass
operators, $\Sigma ^{(k)}(x_1,x_2),$ defined on the
$\tau$-contour. To this aim, let us write the motion equations
for the real-time Green's functions,
\begin{equation}  \label{270}
\begin{array}{rcl}
[i\gamma \partial (x_1)-m]\ G(x_1,x_2) & = & \delta _c
(x_1,x_2)+\int _c d^4 x^{\prime}\ \Sigma(x_1,x^{\prime})\
G(x^{\prime},x_2) \mbox{ ,}
\vphantom{\biggl ]} \\ G(x_1,x_2)\ %
[-i\gamma\stackrel{\leftarrow}{\partial}(x_2) -m] & = &
\delta_c (x_1,x_2)+\int _c d^4 x^{\prime}\ G(x_1,x^{\prime})\
\Sigma(x^{\prime},x_2) \mbox{,}\vphantom{\biggl ]}
\end{array}\end{equation}
The integration in the right-hand side of Eqs. (\ref{270}) is
performed according to the rules:
\begin{equation}  \label{290}
\int _c d^4 x...=\int d\sigma (x\!\mid\! u)\int _c d\tau
....\mbox{ ,}
\end{equation}
and
\begin{equation}  \label{300}
\int _c d\tau ...=\int _{t_0}^\infty d\tau ...\mid _{c^+} -
\int _{t_0}^{\infty} d\tau ... \mid _{c^-} \mbox{ ,}
\end{equation}
where $c^+ (c^-)$ is the upper (lower) branch of the
$\tau$-contour (see Fig. 1). Finally, the delta function on
the $\tau$-contour is defined as
\begin{equation}  \label{310}
\delta _c (x_1,x_2)=\delta _\sigma (x_1,x_2)\delta _c (\tau _1 -\tau _2) %
\mbox{ ,}
\end{equation}
where
\begin{equation}  \label{320}
\delta _c (\tau _1 -\tau _2)=\left\{
\begin{array}{rcc}
\delta (\tau_1 -\tau_2) & \mbox{ if } & \tau_1\in c^+\mbox{
and  }\tau_2\in c^+ \mbox{ ,}\vphantom{\biggl ]} \\ -\delta
(\tau_1 -\tau_2) & \mbox{ if } & \tau_1\in c^- \mbox{  and  }
\tau_2\in c^- \mbox{ ,}\vphantom{\biggl ]} \\ 0 &  & \mbox{
otherwise }.\vphantom{\biggl ]}
\end{array}
\right.
\end{equation}
$\delta_\sigma (x)$ is the $d\!=\!3$ delta function of the
space arguments on the hyperplane $\sigma (x\!\mid\! u)$.

For purpose of comparison with Eq. (\ref{260}), let us write the motion
equations for the correlation function, $G^<(x_1,x_2)$. From Eqs. (\ref{270}%
), the Kadanoff-Baym type equations follow
\begin{equation}  \label{330}
[i\gamma \partial (x_1)-m-\Sigma_{MF}(x_1)]\ G^<(x_1,x_2)=\int
dx^{\prime}\{ \Sigma^<(x_1,x^{\prime})\ G^A(x^{\prime},x_2)
+\Sigma^R(x_1,x^{\prime})\ G^<(x^{\prime},x_2)\}\mbox{ ,}
\end{equation}
\begin{equation}  \label{340}
G^<(x_1,x_2)\ [-i\gamma \stackrel{\leftarrow}{\partial}(x_2)-m-%
\ \Sigma_{MF}(x_2)]=\int dx^{\prime}\{ \ G^<(x_1,x^{\prime})\
\Sigma^A(x^{\prime},x_2) + G^R(x_1,x^{\prime})\
\Sigma^<(x^{\prime},x_2)\}\mbox{ .}
\end{equation}
Here, the mean field part of the mass operator was extracted
according to the following prescription:
\begin{equation}  \label{350}
\begin{array}{rcl}
\Sigma(x_1,x_2) & = & \Sigma_{MF}(x_1)\delta (x_1 -x_2)+
\vphantom{\biggl ]}
\\
& + &
\theta(\tau_1-\tau_2)\Sigma^>(x_1,x_2)+\theta(\tau_2-\tau_1)
\Sigma^<(x_1,x_2)\mbox{ .}\vphantom{\biggl ]}
\end{array}
\end{equation}
Moreover, the retarded and advanced Green's functions were
introduced,
\begin{equation}  \label{360}
\begin{array}{lcl}
G^{R/A}(x_1,x_2) & = & \pm \theta[\pm (\tau_1-\tau_2)%
]\{G^>(x_1,x_2)-G^<(x_1,x_2) \}=\vphantom{\biggl ]} \\
& = & G^c(x_1,x_2)-G^{</>}(x_1,x_2)=G^{>/<}(x_1,x_2)-G^a (x_1,x_2)\mbox{ .}%
\vphantom{\biggl ]}
\end{array}
\end{equation}
In a similar way, the retarded and advanced components of the
mass operator are defined. A mutual exchange of the symbols
$>\,\mbox{ and }\,<$ transforms Eqs. (\ref{330}) and
(\ref{340}) into the corresponding equations for the
correlations functions, $G^> (x_1,x_2)$. By this way, the
Kadanoff-Baym equations (\ref{330}) and (\ref{340}) have the
usual form however they are defined using covariant modified
constituents (Green's functions and mass operators). However,
these small corrections are essential for the search of the
agreement with RKE (\ref{260}).

The transition to the mixed Wigner's representation for the
motion Eqs. (\ref {330}) and (\ref{340}) is usually done with
a limitation to the lowest orders of the gradient
decomposition (the first order into the drift part of RKE and
order zero for the collision part) \cite{KB}, approximation
that is justified only for rather slow kinetic processes. This
allows us to write the motion equations in the simplest local
form. This restriction casts some doubt upon the description
of non-equilibrium processes in nuclear matter at extreme
conditions (see, e.g., \cite{CP,K}). However, for simplicity's
sake, we keep here this approximation only. Then, the motion Eqs. (\ref{330}) 
and (\ref{340}) in the Wigner representation are transformed into
the following equations:
\begin{eqnarray}  \label{370}
[\hat p +\frac i2 \gamma\partial (x)-m-\Sigma_{MF}(x)+\frac
i2\partial_\mu (x)\Sigma_{MF}(x) \partial^\mu (p)]G^< (xp)=
\nonumber \\ =\Sigma^< (x)G^A (xp)+\Sigma^R (x)G^< (xp)\mbox{
,}
\end{eqnarray}
\begin{eqnarray}  \label{380}
G^< (xp)[\hat p -\frac i2 \gamma\stackrel{\leftarrow}{\partial}%
(x)-m-\Sigma_{MF}(x) -\frac i2 \stackrel{\leftarrow}{\partial}%
_\mu\!(p)\partial^\mu (x)\Sigma_{MF}(x)]=  \nonumber \\ =G^<
(xp)\Sigma^A (xp)+G^R (xp)\Sigma^< (xp)\mbox{ .}
\end{eqnarray}
We have neglected gradient terms in the right-hand sides of
these equations \cite{KB}.

\subsection{Generalized RKE}

Let us fulfill now the concordance of the ''generalized RKE''
(\ref{140}) and the system of equations of the Kadanoff-Baym
type (\ref{370}) and (\ref {380}). Obviously, for this aim, it
is necessary to find such a combination of these equations
which provides a coincidence with the left-hand side of Eq.
(\ref{260}) describing the free evolution of the fermion
subsystem. This assumption will permit us to identify (in the
corresponding orders of the gradient expansion) also the other
parts of Eq. (\ref{260}) and Eqs. (\ref {370}), (\ref{380}).

In order to perform such a construction, it is sufficient to multiply Eq. (%
\ref{380}) by $i\hat p$ from the right and Eq. (\ref{370})
from the left and to subtract the second result from the first
one, i.e., it is necessary to consider the following algebraic
combination:
\begin{equation}  \label{390}
Eq.(\ref{380})\,i\hat p -i\hat p \,Eq.(\ref{370}) \mbox{ .}
\end{equation}
Then, we get the following equation
\begin{eqnarray}  \label{400}
p^\mu \partial_\mu (x)G^<(xp)+\frac 12 [\hat p \gamma_\mu
,\partial^\mu (x)G^<(xp)]+ im[\hat p ,G^<(xp)]-  \nonumber \\
-i\{G^<(xp)\Sigma_{MF}(x)\hat p -\hat p \Sigma_{MF}(x)
G^<(xp)\}+ \\ +\frac 12\{\partial_\mu (p)G^<(xp)\partial^\mu
(x)\Sigma_{MF}(x)\hat p+ \hat p\partial_\mu
(x)\Sigma_{MF}(x)\partial^\mu (p)G^<(xp)\}=-S(xp) \mbox{ ,}
\nonumber
\end{eqnarray}
where $S(xp)$ is the collision integral,
\begin{eqnarray}  \label{410}
S(xp)=i\hat p\{\Sigma^<(xp)G^A(xp)+\Sigma^R(xp)G^<(xp)\}-
\nonumber \\ -i\{G^<(xp)\Sigma^A(xp)+
G^R(xp)\Sigma^<(xp)\}\hat p \mbox{ .}
\end{eqnarray}
Now, from a comparison of Eqs. (\ref{260}) and (\ref{400}), it
follows an equivalence of the left and right parts of those
relations:
\begin{eqnarray}  \label{420}
(2\pi)^{-4}\sqrt{p^2}\int dy e^{-ipy}
<[\bar\psi(x_+),H_{MF}(\tau_+)]\psi(x_-)+\bar\psi(x_+)[\psi(x_-),
H_{MF}(\tau_-)]> =  \nonumber \\ =i\{G^<(xp)\Sigma_{MF}(x)\hat
p -\hat p \Sigma_{MF}(x)G^<(xp)\}- \\ -\frac 12\{\partial_\mu
(p)G^<(xp)\partial^\mu (x)\Sigma_{MF}(x)\hat p+ \hat
p\partial_\mu (x)\Sigma_{MF}(x)\partial^\mu (p)G^<(xp)\}\mbox{
,}  \nonumber
\end{eqnarray}
\begin{equation}  \label{430}
S(xp)=-(2\pi)^{-4}\sqrt{p^2}\int dy e^{-ipy}
<[\bar\psi(x_+),H_{r}(\tau_+)]\psi(x_-)+\bar\psi(x_+)[\psi(x_-),
H_{r}(\tau_-)]> \mbox{.}
\end{equation}
Let us remind that these equalities are correct only in the
lowest orders of the gradient expansions.

Both parts of the equality (\ref{420}) describe the force
contribution (with the inverse signs) of the mean fields into
the convective part of RKE (\ref {260}) or (\ref{400}). Let us
illustrate this aspect by means of an actual model. We thus
refer to the standard Walecka model, where the Hamiltonian for
the interaction of nucleons with massive vector and scalar
meson fields reads as \cite{SW}:
\begin{equation}  \label{440}
H_{in}(\tau)=\int d\sigma(x\!\mid\! u)\bar \psi
(x)\{g_v\,\omega^\mu (x)\gamma_\mu - g_s \phi (x)\}\psi (x)
\mbox{ .}
\end{equation}
In the mean field approximation, the relevant Hamiltonian,
$H_{MF}(\tau),$ can be deduced from this one by replacing the
meson fields operators with their mean values, $\phi\to
<\phi>,$ and $\omega^\mu\to <\omega^\mu >$. Then, the
corresponding contribution to the mass operator is equal to
\begin{equation}  \label{450}
\Sigma_{MF}(x)=g_v\!<\!\omega^\mu (x)\!>\!\gamma_\mu -g_s
<\phi (x)> \mbox{.}
\end{equation}
In the first order of the gradient expansion, both RKE
(\ref{260}) and (\ref {400}) lead to identical results in the
frame of the mean field approximation:
\begin{equation}  \label{460}
\begin{array}{lcl}
P^\mu \partial_\mu (x)G^<(xP)+\frac 12\partial^\mu
(x)M(x)\{\hat P,\partial_\mu (P)G^<(xP)\} & + &
\vphantom{\biggl ]} \\ +g_v P^\mu F_{\mu\nu} (x)\partial^\nu
(P)G^<(xP)+\frac12 [\hat P\gamma^\mu ,\partial_\mu (x)G^<(xP)]
& + & \vphantom{\biggl ]} \\ +iM(x)[\hat P,G^<(xP)]-\frac 12
g_v F_{\mu\nu}(x)[\hat P\gamma^\mu ,\partial^\nu (P)G^<(xP)] &
= & -S(xP) \mbox{ ,} \vphantom{\biggl ]}
\end{array}
\end{equation}
where $M(x)=m-g_s\!<\!\phi (x)\!>\!$ is the effective nucleon
mass in the mean field approximation,
$F^{\mu\nu}(x)=\partial^\mu \!<\!\omega^\nu\!>\!-
\partial^\nu \!<\!\omega^\mu\!>\!$ and $P^\mu =p^\mu
-g_v\!<\!\omega^\mu(x)\!>\!$ is the kinetic momentum.

RKE (\ref{460}) converts in RKE of the Vlasov type when
conditions of the collision effects are neglected $(S(xP)=0)$.
The Vlasov RKE in this form was obtained and analyzed for the
first time in Refs. \cite{S3}-\cite{S5}. These equations are
rather complicated because the spin and meson degrees of
freedom as well as states with positive and negative energies
are taken into account. On the basis of this RKE, a series of
more simple, particular cases can be analyzed. The simplest
situation corresponds to the case of a spin saturated system
without antinucleon states, where all the spin-dependent
effects in the RKE can be neglected (the spin effects were
investigated separately in Ref.~\cite{S3}). Using the Clifford
decomposition of the Wigner function, we get -- in the
quasi-classical limit -- the well-known RKE of the Vlasov type
in the quantum hadrodynamics,
\begin{eqnarray}  \label{470}
P^\mu\partial_\mu (x)f^S(xP)+M\partial^\mu(x)M\partial_\mu
(P)f^S(xP)+g_v P_\mu F^{\mu\nu}\partial_\nu (P)f^S(xP)=0
\mbox{ ,}
\end{eqnarray}
where $f^S(xP)$ corresponds to the scalar part of the Clifford
decomposition of function $G^<(xP)$ (see Eq. (\ref{190})). It
is important that the basic RKE (\ref{460}) leads to a correct
RKE (\ref{470}) of the Vlasov type for a spin saturated system
without any additional assumption, like that of the rescaling
procedure type (see the detailed discussion in \cite{S3}).

Technical details of derivation of Eqs.(\ref{460})-(\ref{470})
are given in Refs.~\cite{S3}-\cite{S5}.
 It is not difficult to obtain analogous RKE also without
 use of the gradient expansion. Such a RKE has a non-local
 character in the momentum space.

Calculation of collision integrals can be performed using two
alternative schemes based either on formula (\ref{410}) or
(\ref{430}). In the latter case, a standard perturbation
theory was developed in Refs.~\cite {S1,S2,S6,S7,S8}. Within
the framework of the Walecka model (\ref{440}), this approach
allows us to derive collision integrals of the Bloch type
(when the meson subsystem is considered on equal terms with
the nucleon one) or of the Boltzmann type (if the role of
mesons is confined only to the formation of the effective
nucleon-nucleon interaction).

Let us return to the representation of the collision integral
(\ref{410}) based on the method of modified real-time Green's
functions. Our subsequent
transformations are based on the analogy with the nonrelativistic theory~%
\cite{Z2,KB}. From the definition (\ref{360}), it follows that
the Green's functions and mass operator satisfy the
identities:
\begin{equation}  \label{480}
\begin{array}{c}
G^R(xp)-G^A(xp)=G^>(xp)-G^<(xp) \mbox{ ,} \vphantom{\biggl ]}
\\
\Sigma^R(xp) -\Sigma^A(xp)=\Sigma^>(xp)-\Sigma^<(xp) \mbox{ .}
\end{array}
\end{equation}
Using these relations, the collision integral (\ref{410}) can
be transformed into the form
\begin{equation}  \label{490}
\begin{array}{lcl}
S(xp) & = & \frac i2\{\hat p [\Sigma^>(xp)G^<(xp)- \Sigma^<(xp)G^>(xp)]+%
\vphantom{\biggl ]} \\ & + & [G^<(xp)\Sigma^>(xp)-
G^>(xp)\Sigma^<(xp)]\hat p\}+S^{off}(xp) \mbox{ ,}
\end{array}
\end{equation}
where $S^{off}(xp)$ is the off-mass-shell part of the
collision integral,
\begin{equation}  \label{500}
\begin{array}{lcl}
S^{off}(xp) & = & \frac i2\{\hat p
\Sigma^<(xp)[G^R(xp)+G^A(xp)]+ \hat p
[\Sigma^R(xp)+\Sigma^A(xp)]G^<(xP)- \vphantom{\biggl ]} \\ & -
& G^<(xp)[\Sigma^R(xp)+\Sigma^A(xp)]\hat p
-[G^R(xp)+G^A(xp)]\Sigma^<(xp)\hat p\} \mbox{ .}
\end{array}
\end{equation}

Of course, the off--shell effects enter in the whole collision
operator (49), but function (50) refers to them exclusively
for the present aims. The separation into two parts, (49) and
(50), is ascertained by means of spectral properties of the
Green's functions and the mass operator. In our formalism, the
corresponding spectral decompositions can be fulfilled in a
covariant form (see Appendix B). This circumstance is a very
desirable property of the relativistic kinetic theory which is
intended for a description of macroscopic motions of a medium.
Let us remark that the off-shell processes play the most
important role in the relativistic nuclear physics of
intermediate \cite{me} and high \cite{he} energies.

The generalized RKE (\ref{400}) with the collision integral (\ref{490}), (%
\ref{500}) are the main results of this Section. It is worth
remarking that a few other versions of generalized RKE have
been previously obtained in the literature (see, e.g.,
\cite{MH} and \cite{MH92}).
 We have in mind the cases of derivation of RKE for
the Fermi subsystem based either on the Dirac type equations
of motion for the field operator or on the equations of the
Kadanoff-Baym type (\ref{330}), (\ref {340}). However, it is
clear that this procedure contains some elements of an
indetermination and arbitrariness on the stage
of the construction of the motion equation for the Wigner function. In 
considered our case, the selection rule (\ref{390}) has a
reliable dynamical basis (see Sec. 2) which was tested in
Refs. \cite{S5}-\cite{S10}. Of course, the generalized RKE
(\ref{400}) can be obtained also by a direct way on the basis
of Eq. (\ref{260}), without using the rule (\ref{390}), with a
help of the Dyson equation and definitions of the type of Eqs.
(\ref{170}).

RKE (\ref{400}), (\ref{490}), and (\ref{500}) have a more
complicated matrix structure. This characteristic behavior
arises by taking into account inner degrees of freedom which
are intrinsic for the considered Fermi subsystem and, in
particular, states with positive and negative energies.
Indeed, the presence of $\gamma ^0$ matrices before the time
derivative in equations of the Kadanoff-Baym type (\ref{330})
and (\ref{340}) ensures the ''direct'' time movement for
states of the spinor fields with positive energies and the
''inverse'' time direction for states with negative energies.
But for the kinetic theory, it is characteristic the existence
of a single time arrow. The rule (\ref{390}) just ensures a
covariant way of projection of Eqs. (\ref {330}) and
(\ref{340}) on such a single time direction.

Finally, let us return once more to the role of the assumption
$p^2>0$ (Assumption 1). A consecutive realization of this
limitation leads to appreciable complication of the considered
integrals. Similar restriction is absent in the traditional
approach where an integration is carried out in the whole
momentum space. In order to get traditional results in the
frame of our formalism, it is necessary to implement the
extension of obtained RKE
into space-like region. There are three aspects of such a continuation:%
\newline
1) into the integrals with respect to the momentum
space,\newline 2) into the time differential
operations,\newline
3) into the time integral operations.

The extension of the momentum integrals into the region,
$p^2<0,$ is trivial in the quasiparticle approximation (Sec.
4).

As for the derivation of the Vlasov RKE, the considered continuation does
not lead to some changes of differential operations 
with respect to the time (e.g., to a change $\partial/\partial\tau$ by $%
i\partial/\partial\tau$). Indeed, in the mean field
approximation, the
calculation of the force part of the Vlasov RKE (the right-hand side in Eq.(%
\ref{260})) results in the cancellation of $\sqrt{p^2}$ once rule (\ref{240}%
) has been used. In the other words, RKE conserves its
physical content.

Finally, in the Markovian approximation, integral time operations do not
lead to any modification of results at the continuation into the region $%
p^2<0$.

The possibility of continuation beyond the above-mentioned
cases needs further researches.

\section{ Quasiparticle approximation}

In this Section, we derive a covariant generalization of the
quasiparticle approximation (QPA), widely used in the kinetic
theory, which allows us to leave, by the simplest way, the
framework of the usual perturbation theory. After that, we
check possibilities of the suggested method on an example of
derivation of collision integral of the
Boltzmann-Uehling-Uhlenbeck (BUU) type for the standard
Walecka model of relativistic nuclear matter consisting of
nucleons, scalar and vector mesons with the interaction
Hamiltonian (\ref{440}). Finally, we discuss briefly a
feasible generalization of the QPA in the model (\ref{440}).

Let us introduce a non-equilibrium spectral function in the
Wigner representation (see the first formula in Eq.$(48)$)
\begin{equation}
a(xp)=i[G^{R}(xp)-G^{A}(xp)]=i[G^{>}(xp)-G^{<}(xp)]\mbox{ ,}
\label{510}
\end{equation}
This function satisfies the following sum rule
\begin{equation}
{(2\pi )}^{3}\int dE\,a(x;p_{\perp },E)=\gamma u\mbox{ ,}
\label{520}
\end{equation}
which is a simple consequence of the single-time
anti-commutation relations for the field operators and of the
rule (\ref{240}). In Eq. (\ref{520}), the integration is
carried out over the longitudinal energy, $E=pn;\;p_{\perp
}^{\mu }=\Delta ^{\mu \nu }p_{\nu }$ is the transverse
momentum ($\Delta ^{\mu \nu }$ is the projection operator
(\ref{90})).


$G^{>}(xp)$ and $G^{<}(xp),$ can be represented as:
\begin{equation}
\begin{array}{lcl}
G^{<}(xp) & = & ia(xp)\mathcal{F}(xp)\vphantom{\biggl ]}\mbox{
,} \\ G^{>}(xp) & = & -ia(xp)[1-\mathcal{F}(xp)]\mbox{ ,}
\end{array}
\label{530}
\end{equation}
where $\mathcal{F}(xp)$ is a unknown function.

To determine the non-equilibrium spectral function, we consider relation (%
\ref{510}) and the corresponding motion equations for the
retarded and advanced Green's functions. We write at once
these equations in the minimal order of the gradient
expansion, namely,
\begin{equation}
\begin{array}{r}
\lbrack \hat{p}+\frac{i}{2}\gamma \partial (x)-m-\Sigma _{MF}(x)+\frac{i}{2}%
\partial _{\mu }(x)\Sigma _{MF}(x)\partial ^{\mu }(p)]G^{R/A}(xp)= \\
=1+\Sigma ^{R/A}(xp)\ G^{R/A}(xp)\,,\vphantom{\biggl ]} \\
G^{R/A}(xp)[\hat{p}-\frac{i}{2}\gamma \stackrel{\leftarrow }{\partial }%
(x)-m-\Sigma _{MF}(x)-\frac{i}{2}\stackrel{\leftarrow
}{\partial }_{\mu }\!(p)\partial ^{\mu }(x)\Sigma _{MF}(x)]=
\\
=1+G^{R/A}(xp)\ \Sigma ^{R/A}(xp)\,.\vphantom{\biggl ]}
\end{array}
\label{540}
\end{equation}
Since the gradient expansion orders in the drift and
collisions parts of RKE (\ref{400}) are fixed, we can restrict
ourselves to order zero of this decomposition in the
estimation of correlation functions (\ref{530}), so that
\cite{MH},
\begin{equation}
\lbrack \hat{p}-m-\Sigma _{MF}(x)-\Sigma ^{R/A}(xp)]\
G^{R/A}(xp)=1\,. \label{550}
\end{equation}
In our toy model, $\Sigma _{MF}(x)$ is defined by Eq.
(\ref{450}) and hence,
\begin{equation}
\lbrack \hat{P}-M(x)-\Sigma ^{R/A}(xp)]\,G^{R/A}(xp)=1\,.
\label{560}
\end{equation}
Let us assume now --- as a working approximation --- that the
mass operator, $\Sigma (xp),$ has the same matrix
structure as the mean field part (\ref{450}) (in the Walecka model (\ref{440}%
), this implies a possibility to neglect the tensor part of
Clifford's decomposition of the mass operator),
\begin{equation}
\Sigma (xp)\cong \Sigma _{s}(xp)+\gamma ^{\mu }\ \Sigma _{\mu
}(xp) \label{570}
\end{equation}
 (a more general form is given in Ref.\cite{H}). In this
expression as well as the following ones, the marks $R/A$ are
omitted for simplicity's sake. Eq. (\ref{560}) has then a
matrix structure like the Dirac equation of noninteracting
fields (quasiparticle structure)
\begin{equation}
\lbrack \hat{\mathcal{P}}(xp)-\mathcal{M}(xp)]\
G^{R/A}(xp)=1\,,  \label{580}
\end{equation}
where
\begin{equation}
\mathcal{P}_{\mu }(xp)=P_{\mu }(xp)-\Sigma _{\mu }(xp)\,,
\label{590}
\end{equation}
\begin{equation}
\mathcal{M}(xp)=M(xp)-\Sigma _{s}(xp)\,.  \label{600}
\end{equation}
Therefore, Eq. (\ref{580}) can be easily solved:
\begin{equation}
G^{R/A}(xp)=\frac{\hat{\mathcal{P}}+\mathcal{M}}{\mathcal{P}^{2}-\mathcal{M}%
^{2}\pm i\epsilon E}\ .  \label{610}
\end{equation}

Then, from Eqs. (\ref{510}) and (\ref{610}), we get
\begin{equation}
a(x\mathcal{P})=2\pi (\hat{\mathcal{P}}+\mathcal{M})\delta [Re(\mathcal{P}%
^{2}-\mathcal{M}^{2})]\{\theta (E)-\theta (-E)\}\ .
\label{620}
\end{equation}
This result is correct, provided that the quasiparticle
excitations are weakly damped, i.e.,
\begin{equation}
|Re(\mathcal{P}^{2}-\mathcal{M}^{2})|>>|Jm(\mathcal{P}^{2}-\mathcal{M}%
^{2})|\ .  \label{630}
\end{equation}
It is easy to verify that the spectral function (\ref{620})
satisfies the sum rule (\ref{520}).


The substitution of the non-equilibrium spectral function
(\ref{620}) into Eqs. (\ref{530}) leads to the following
expressions for the correlation functions,
\begin{equation}  \label{640}
\begin{array}{lcl}
G^<(x\mathcal{P}) & = & 2\pi i(\hat {\mathcal{P}} +\mathcal{M})\delta (%
\mathcal{P^{\prime}}^2-\mathcal{M^{\prime}}^2)\mathcal{F}(x\mathcal{%
P^{\prime}})\, ,\vphantom{\biggl ]} \\
G^>(x\mathcal{P}) & = & -2\pi i(\hat {\mathcal{P}} +\mathcal{M}) \delta (%
\mathcal{P^{\prime}}^2-\mathcal{M^{\prime}}^2)[1- \mathcal{F}(x\mathcal{%
P^{\prime}})] \ ,
\end{array}
\end{equation}
where $\mathcal{P^{\prime}} =Re \, \mathcal{P}$ and
$\mathcal{M^{\prime}} =Re \, \mathcal{M}$.


As an illustration, we present a derivation of a collision
integral of the BUU type in the Walecka model (\ref{440}).
This example is well known in the literature, therefore it is
convenient for testing our method. We confined ourselves to
the Born approximation only in the calculation of the self
energy parts, $\Sigma ^{>/<},$ in the collision integral
(\ref{490}). The corresponding diagrams are shown in Fig.~2.

\begin{picture}(160,70)(10,5)
\thicklines 
\multiput(20,44)(0,2){10}{\line(0,1){1.5}}
\multiput(40,44)(0,2){10}{\line(0,1){1.5}}
\put(20,44){\line(1,0){20}} \qbezier(20,64)(30,75)(40,64)
\qbezier(20,64)(30,55)(40,64) \put(16,43){${\scs +}$}
\put(16,65){${\scs +}$} \put(41,65){${\scs -}$}
\put(41,43){${\scs -}$} \put(20,55){\vector(0,1){2}}
\put(40,55){\vector(0,-1){2}} \put(29,44){\vector(1,0){1}}
\put(31,69.5){\vector(-1,0){1}} \put(29,59.5){\vector(1,0){1}}
\put(50,0){\begin{picture}(35,40)
\multiput(0,0)(0,8){2}{\qbezier(20,44)(19,46)(20,48)
\qbezier(20,48)(21,50)(20,52)} \qbezier(20,60)(19,62)(20,64)
\multiput(0,0)(0,8){2}{\qbezier(40,44)(41,46)(40,48)
\qbezier(40,48)(39,50)(40,52)} \qbezier(40,60)(41,62)(40,64)
\put(20,44){\line(1,0){20}} \qbezier(20,64)(30,75)(40,64)
\qbezier(20,64)(30,55)(40,64) \put(16,43){${\scs +}$} \put(16,65){${\scs +}$%
} \put(41,65){${\scs -}$} \put(41,43){${\scs -}$}
\put(19.5,53.5){\vector(0,1){2}}
\put(40.5,55){\vector(0,-1){2}} \put(29,44){\vector(1,0){1}}
\put(31,69.5){\vector(-1,0){1}}
\put(29,59.5){\vector(1,0){1}} \end{picture}} 
\put(100,0){\begin{picture}(35,40)
\multiput(20,44)(0,2){10}{\line(0,1){1.5}}
\multiput(40,44)(0,2){10}{\line(0,1){1.5}}
\put(20,64){\line(1,0){20}} \put(20,64){\line(1,-1){20}}
\put(20,44){\line(1,1){20}}
\put(20,64){\vector(1,-1){16}} \put(20,44){\vector(1,1){5}} \put(16,43){$%
{\scs +}$} \put(16,65){${\scs +}$} \put(41,65){${\scs -}$} \put(41,43){$%
{\scs -}$} \put(20,55){\vector(0,1){2}}
\put(40,55){\vector(0,-1){2}}
\put(29,64){\vector(-1,0){1}} \end{picture}} 
\put(0,-35){\begin{picture}(35,40)
\multiput(0,0)(0,8){2}{\qbezier(20,44)(19,46)(20,48)
\qbezier(20,48)(21,50)(20,52)} \qbezier(20,60)(19,62)(20,64)
\multiput(0,0)(0,8){2}{\qbezier(40,44)(41,46)(40,48)
\qbezier(40,48)(39,50)(40,52)} \qbezier(40,60)(41,62)(40,64)
\put(20,64){\line(1,0){20}} \put(20,64){\line(1,-1){20}}
\put(20,44){\line(1,1){20}} \put(20,64){\vector(1,-1){16}}
\put(20,44){\vector(1,1){5}} \put(16,43){${\scs +}$}
\put(16,65){${\scs +}$} \put(41,65){${\scs -}$}
\put(41,43){${\scs -}$} \put(19.5,53.5){\vector(0,1){2}}
\put(40.5,55){\vector(0,-1){2}}
\put(29,64){\vector(-1,0){1}} \end{picture}} 
\put(50,-35){\begin{picture}(35,40)
\multiput(0,0)(0,8){2}{\qbezier(20,44)(19,46)(20,48)
\qbezier(20,48)(21,50)(20,52)} \qbezier(20,60)(19,62)(20,64)
\multiput(40,44)(0,2){10}{\line(0,1){1.5}}
\put(20,64){\line(1,0){20}} \put(20,64){\line(1,-1){20}}
\put(20,44){\line(1,1){20}}
\put(20,64){\vector(1,-1){16}} \put(20,44){\vector(1,1){5}} \put(16,43){$%
{\scs +}$} \put(16,65){${\scs +}$} \put(41,65){${\scs -}$} \put(41,43){$%
{\scs -}$} \put(19.5,53.5){\vector(0,1){2}}
\put(40,55){\vector(0,-1){2}}
\put(29,64){\vector(-1,0){1}} \end{picture}} 
\put(100,-35){\begin{picture}(35,40)
\multiput(20,44)(0,2){10}{\line(0,1){1.5}}
\multiput(0,0)(0,8){2}{\qbezier(40,44)(41,46)(40,48)
\qbezier(40,48)(39,50)(40,52)} \qbezier(40,60)(41,62)(40,64)
\put(20,64){\line(1,0){20}} \put(20,64){\line(1,-1){20}}
\put(20,44){\line(1,1){20}} \put(20,64){\vector(1,-1){16}}
\put(20,44){\vector(1,1){5}} \put(16,43){${\scs +}$}
\put(16,65){${\scs +}$} \put(41,65){${\scs -}$}
\put(41,43){${\scs -}$} \put(20,55){\vector(0,1){2}}
\put(29,64){\vector(-1,0){1}} \put(40.5,54.5){\vector(0,-1){1}} \end{picture}%
} \end{picture}

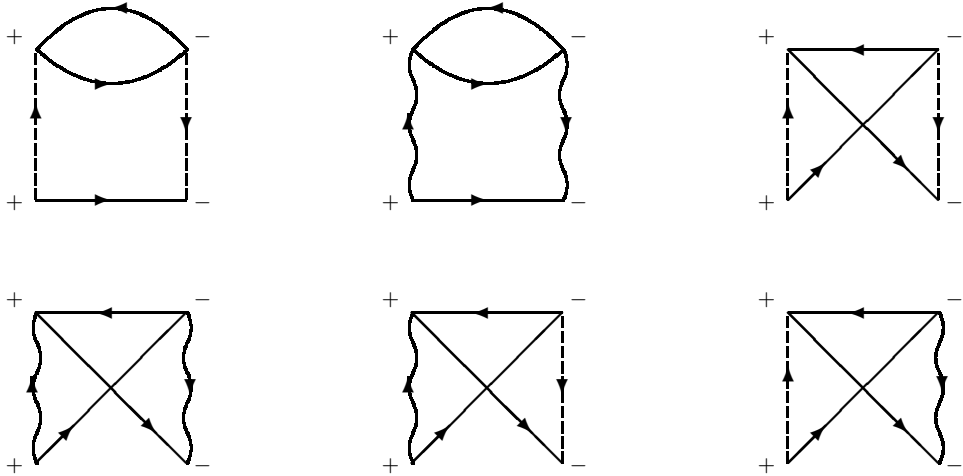
Fig. 2. Born diagrams for the self-energy part, $-i\Sigma^<=
-i\Sigma^{-+}$. The solid, dashed and wavy lines denote the
propagators of nucleons, scalar and vector mesons,
respectively.

\vspace*{0.5cm}

In a general case, this collision integral has a rather
complicated form since we take into account simultaneously the
spin and meson degrees of freedom and states with positive and
negative energies. The simplest situation corresponds to a
spin saturated nuclear matter without antinuclear component.
In order to get the corresponding collision integral, we have
to perform a transform from the spinor representation to the
spin one for states with positive energies \cite{G} and to
take into account only the diagonal part of the Wigner
function in the spin representation. As a result, for function
$\mathcal{F}(x\mathcal{P})$ in Eq. (\ref{640}), we
obtain (for simplicity, we omit here and below all primes in functions $%
\mathcal{M}^{\prime}\mbox{ and } \mathcal{P}^{\prime}$)
\begin{equation}  \label{650}
\mathcal{F}_{\alpha\beta}(x\mathcal{P})\Rightarrow
\sum_{r,s=1,2}u^s_\alpha
\bar u^r_\beta \Phi_{sr}(x\mathcal{P})\Rightarrow\frac{ (\hat {\mathcal{P}}+%
\mathcal{M})_ {\alpha\beta}}{2\mathcal{P}^0}\,\Phi
(x\mathcal{P}) \, ,
\end{equation}
where $\Phi_{sr} (x\mathcal{P})$ is the  Wigner function of
states with positive energy in the spin representation. For a
spin-saturated nucleon subsystem we have $\Phi_{rs}\simeq
\Phi\delta_{rs}$, where $\Phi (x\mathcal{P})$ is the scalar
Wigner function. In Eq.(\ref{650}), we used the completeness
relation for the Dirac spinor
\cite{BD} \[ \sum_{r=1,2}u^r\bar u^r =\frac{\hat %
{\mathcal{P}}+\mathcal{M}}{2\sqrt{%
\mathcal{M}^2+ \bar {\mathcal{P}}^2}}\, .  \] Substitution of
Eq. (\ref{650}) into the first formula of Eqs. $(64)$ leads to
the relation
\begin{equation}  \label{660} G^{<(+)}(x\mathcal{P})
=i\pi \frac{\hat {\mathcal{P}}+\mathcal{M}}{\mathcal{P%
 }^0} \delta \biggl(
\mathcal{P}^0 - \sqrt{\mathcal{M}^2+ \bar {\mathcal{P}}^2
}\biggr) F(x\mathcal{P}) \, , \end{equation} where a new
auxiliary function is introduced:  \begin{equation}
\label{670}
F(x\mathcal{P})=\frac{\mathcal{M}}{\mathcal{P}^0}\Phi
(x\mathcal{P})\ .
\end{equation}
This expression differs from the true Wigner function, $\Phi
(x\mathcal{P}),$ by the scaling factor,
$\mathcal{M}/\mathcal{P}^0$ \cite{Elze87}. Introduction of
such a rescaled Wigner function (\ref{670}) is convenient in
order to write the collision integral.

A tedious but straightforward evaluation of the on-mass-shell
part of the collision integral (\ref{490}) leads to the
following result:
\begin{eqnarray}  \label{680}
S^{(+)}(x\mathcal{P})=(2\pi)^{-5} \int\prod_{i=2}^4 \frac{d^3\mathcal{P}_i}{%
\sqrt{\mathcal{M}^2+ \bar {\mathcal{P}}^2_i}} W(\mathcal{P}, \mathcal{P}_2,%
\mathcal{P}_3, \mathcal{P}_4) \delta (\mathcal{P}-\mathcal{P}_2-\mathcal{P}%
_3-\mathcal{P}_4)  \nonumber \\
\{F(x\mathcal{P})F(x\mathcal{P}_2)\bar F(x\mathcal{P}_3)\bar F(x\mathcal{P}%
_4) -\bar F(x\mathcal{P})\bar F(x\mathcal{P}_2) F(x\mathcal{P}_3) F(x%
\mathcal{P}_4)\}\, ,
\end{eqnarray}
where $\bar F=1-F$ is the Pauli blocking factor. Here, the
transition rates are defined as follows:
\begin{equation}  \label{690}
W=W_{ss}+W_{vv}+W_{sv}\, .
\end{equation}
The partial transition rates are given by exchange of the
corresponding pairs of the scalar or vector mesons:
\begin{eqnarray}  \label{700}
W_{ss}(\mathcal{P}\mathcal{P}_2\mathcal{P}_3\mathcal{P}_4)&=&4g_s^4\{4D_s^2(%
\mathcal{P}-\mathcal{P}_3) [\mathcal{M}^4+\mathcal{M}^2(\mathcal{P}\mathcal{P%
}_3+\mathcal{P}_2\mathcal{P}_4) +  \nonumber \\
&&+(\mathcal{P}\mathcal{P}_3)(\mathcal{P}_2\mathcal{P}_4)]-D_s(\mathcal{P}-%
\mathcal{P}_3)D_s(\mathcal{P} -\mathcal{P}_4)[\mathcal{M}^4+
\nonumber \\
&&+\mathcal{M}^2(\mathcal{P}\mathcal{P}_2+\mathcal{P}\mathcal{P}_3+\mathcal{P%
}\mathcal{P}_4+\mathcal{P}_2\mathcal{P}_3+\mathcal{P}_2\mathcal{P}_4+%
\mathcal{P}_3\mathcal{P}_4)+  \nonumber \\
&&+(\mathcal{P}_2\mathcal{P}_4)(\mathcal{P}\mathcal{P}_3)+(\mathcal{P}_2%
\mathcal{P}_3)(\mathcal{P}\mathcal{P}_4)-(\mathcal{P}\mathcal{P}_2)(\mathcal{%
P}_3\mathcal{P}_4)]\}\mbox{ ,}  \nonumber \\
W_{vv}(\mathcal{P}\mathcal{P}_2\mathcal{P}_3\mathcal{P}_4)&=&8g_v^4 \{4D_v^2(%
\mathcal{P}-\mathcal{P}_3)[2\mathcal{M}^4-\mathcal{M}^2(\mathcal{P}_2%
\mathcal{P}_4+\mathcal{P}\mathcal{P}_3)+ \\
&&+(\mathcal{P}\mathcal{P}_2)(\mathcal{P}_3\mathcal{P}_4) +(\mathcal{P}%
\mathcal{P}_4)(\mathcal{P}_2\mathcal{P}_3)]-  \nonumber \\
&&-D_v(\mathcal{P}-\mathcal{P}_3)D_v(\mathcal{P}-\mathcal{P}_4)[-2\mathcal{M}%
^4+\mathcal{M}^2(\mathcal{P}_2\mathcal{P}_3+\mathcal{P}\mathcal{P}_4+
\nonumber \\
&&+\mathcal{P}\mathcal{P}_3+\mathcal{P}\mathcal{P}_2+\mathcal{P}_3\mathcal{P}%
_4+\mathcal{P}_2\mathcal{P}_4)-2(\mathcal{P}\mathcal{P}_2)(\mathcal{P}_3%
\mathcal{P}_4)]\}\mbox{ ,}  \nonumber \\
W_{sv}(\mathcal{P}\mathcal{P}_2\mathcal{P}_3\mathcal{P}_4)&=&8
g_s^2 g_v^2
\{-4D_s(\mathcal{P}-\mathcal{P}_3)D_v(\mathcal{P}-\mathcal{P}_3)\mathcal{M}%
^2[\mathcal{P}\mathcal{P}_2+  \nonumber \\
&&+\mathcal{P}\mathcal{P}_4+\mathcal{P}_2\mathcal{P}_3+\mathcal{P}_3\mathcal{%
P}_4]+D_s(\mathcal{P}-\mathcal{P}_3)D_v(\mathcal{P}-\mathcal{P}_4)[2\mathcal{%
M}^4+  \nonumber \\
&&+\mathcal{M}^2(-\mathcal{P}\mathcal{P}_2+2\mathcal{P}\mathcal{P}_3-%
\mathcal{P}\mathcal{P}_4+2\mathcal{P}_2\mathcal{P}_4-\mathcal{P}_2\mathcal{P}%
_3-  \nonumber \\
&&-\mathcal{P}_3\mathcal{P}_4)+2(\mathcal{P}\mathcal{P}_3)(\mathcal{P}_2%
\mathcal{P}_4)]\}\mbox{ .}  \nonumber
\end{eqnarray}
Finally, $D_{s,v}(\mathcal{P})$ are the Fourier transforms of
the mesons Green's functions,
 $
D_{s,v}(\mathcal{P})=(\mathcal{P}^2-m_{s,v}^2)^{-1} \mbox{ ,}
$
and $m_s$ and $m_v$ are masses of the scalar and vector mesons, respectively.

It is worth noticing that similar results have been recently
obtained in Ref. \cite{SC} within the framework of
non-covariant real-time Green's functions method. A comparison
of our results with those of Ref. \cite{SC} shows an overall
good agreement (up to a replacement of the fermionic Green
function calculated in the mean--field approximation by that
obtained in the more general approximation (64)). However, it
needs to keep in mind that the standard Walecka model is too
primitive for a proper description of real properties of
nuclear matter at intermediate energies. Already at an
equilibrium state, it leads to an equation of state stiffer
than the expected one and, at moderately high density and
temperature, the effective masses of nucleons become very
small or even negative \cite{ZM}. At present, two directions
of the Walecka model improvement exist: either an increase of
the number of its constituents \cite
{BF} or a transition to a nonlinear generalization of the model \cite{ZM,DC}.
The first way leads to technical complications of the
kinetic theory (by virtue of an essential increase of the
number of constituents for real system), the second one
remains still unexplored for a description of non-equilibrium
states.

\section{Connection with the non-equilibrium statistical operator method}

As previously outlined, in the definition of Wigner's
(\ref{10}) and Green's functions (\ref{170}), the statistical
averaging was performed by means of the equilibrium density
matrix, $\rho,$ in the Heisenberg picture corresponding
ordinarily to a system state at an infinite past. This fact
limits the applicability of the theory to the case of weakly
non-equilibrium states which can be described defining only
slowly varying thermodynamic functions of the system such as
temperature, chemical potential, etc. Now, let us discuss
briefly a possibility of a generalization of the formalism
suggested in Sec. 3 which is based on the change of the
statistical averaging procedure, using the non-equilibrium
statistical operator, $\rho (t) $. It is expected that such a
generalization will allow us to describe strongly
non-equilibrium states of a system \cite{Z2,KB,ML,Z1}. The
non-equilibrium statistical operator method was adapted to
fulfill requirements of the relativistic kinetic theory in our
previous works~\cite {S1,S2,S6,S7}. Here, we discuss only a
problem where this method is combined with the covariant
formalism of real-time Green's functions method (Sec. 3).

Let us write the motion equation for the non-equilibrium
statistical operator in a differential form \cite{S2}
\begin{equation}  \label{710}
d\rho (\tau)/d\tau = -\epsilon \{\rho (\tau)-\rho _q
(\tau)\}\mbox{ ,}
\end{equation}
and, in an integral form ($\epsilon$ is an infinitely small
value, $\epsilon
>0$, and will approach to zero after the execution of the thermodynamic
limiting transition),
\begin{eqnarray}  \label{720}
\rho (\tau)={\rho}_q(\tau)+i\int\limits^{\tau}_{- \infty} d
\tau ^{\prime}e^{\epsilon (\tau ^{\prime}-\tau )} \ \{ [
{\rho}_q (\tau ^{\prime}), H(\tau ^{\prime})] + i d {\rho}_q (
\tau ^{\prime})/ d \tau ^{\prime}\} \mbox{ .}
\end{eqnarray}
Here, the time, $\tau,$ is defined by Eq. (\ref{60}) with a
subsequent introduction of Assumptiones 1 and 2 and relation
(\ref{120}). An infinitely small source in the right-hand side
of Eq. (\ref{710}) is introduced to break the symmetry of this
equation with respect to the time reflection (in the Wigner
sense). The form of Eq. (\ref{710}) allows us to select the
retarded solution of the Liouville equation on the basis of an
analogy with the formal scattering theory. This implies that
after completing our
calculations the thermodynamical limit should be taken and, then, the limit $%
\epsilon\to 0$ assumed.

The quasi-equilibrium statistical operator, $\rho_q (\tau),$
in Eq. (\ref {710}) is the asymptotic form of the
non-equilibrium statistical operator, $\rho (\tau),$ at the
kinetic stage of evolution, when $\tau \to -\infty$. An
explicit form of operator $\rho_q (\tau)$ can be derived from
the principle of the maximum of the information entropy under
a supplementary condition of
the given averaging value of $<P_{\alpha\beta}(xy)>_{q\tau}$, where $%
<...>_{q\tau}=Tr...\rho_q(\tau)$ and the operator,
$P_{\alpha\beta}(xy),$ is defined by Eq. (\ref{20}). For our
task, it is convenient to associate the proper time, $\tau,$
with the slow variable $x^{\mu}$ according to the expression
(\ref{60}).

As a result, we have \cite{S2}
\begin{equation}  \label{730}
{\rho}_q(\tau ) = exp \{ - S(\tau ) \} \mbox{ ,}
\end{equation}
where $S(\tau)$ is the entropy operator of the system at the
kinetic stage,
\begin{equation}  \label{740}
S(\tau ) = \Phi (\tau )+\int d \sigma (x|u) \ \int d^4 y \
P_{\alpha \beta} (x,y) \ F_{\alpha \beta} (x,y),
\end{equation}
and $\Phi(\tau)$ is the normalizing functional defined by
condition $Tr\ \rho_q(\tau)=1$. The Lagrange factor,
$F_{\alpha \beta} (x,y),$ is determined by the
self-consistency condition
\begin{equation}  \label{750}
<P_{\alpha\beta}(xy)>_\tau
\,=\,<P_{\alpha\beta}(xy)>_{q\tau}\mbox{ .}
\end{equation}
In the left-hand side of this equality, the averaging is
performed by means
of the non-equilibrium statistical operator $\rho (\tau)$, i.e., $%
<...>_\tau=Tr...\rho (\tau)$. Relation (\ref{750}) shows a
full equivalence of both statistical operators, $\rho (\tau)$
and $\rho_q (\tau),$ at the kinetic stage of evolution. Hence,
both variants of the Wigner function,
\begin{equation}  \label{760}
f_{\alpha \beta }(x,p)= (2\pi )^{-4}\int dy\,
e^{-ipy}<P_{\alpha \beta }(x,y)>_\tau = (2\pi )^{-4}\int dy\,
e^{-ipy}<P_{\alpha \beta }(x,y)>_{q \tau} \mbox{ ,}
\end{equation}
can be used for a kinetic description of both
quasi-equilibrium and non-equilibrium states.

The Green's functions, (\ref{160})-(\ref{280}) and
(\ref{360}), also permit an analogous generalization for the
case of strong non-equilibrium states.

\begin{equation}  \label{770}
G_{(non-eq)}(xp)=-i(2\pi)^{-4}\int dy e^{ipy}<T_c[\psi
(x_+)\bar \psi (x_-)]>_{ \tau}\, .
\end{equation}
The form of the Kadanoff-Baym Eqs. (\ref{330}), (\ref{340})
remains also the same, taking into account that the
$\epsilon$-terms generated by the right-hand of Eq.
(\ref{720}) vanish after a realization of the thermodynamical
limit. However, a restriction to the lowest terms of the
gradient decomposition is incompatible with a description of
strong non-equilibrium states. Therefore, the corresponding
results of Sec. 4 require a further generalization. This field
of relativistic kinetic theory is still scarcely investigated.

Let us remark that Eq. (\ref{720}) can be rewritten in a form
convenient for the construction of an iteration scheme under
the energy of particle-particle interactions. In full analogy
with the non-relativistic case \cite{Z2}, we obtain
\cite{S1,S2}
\begin{eqnarray}  \label{780}
&\rho ( \tau ) ={\rho}_q( \tau ) + i \int\limits^{\tau}_{-
\infty} d\tau ^{\prime}e^{\epsilon (\tau ^{\prime}-\tau)} \{ \
[ \ {\rho}_q ( \tau ^{\prime}), H_{in}(\tau ^{\prime}) \ ] + &
\nonumber \\ &+\int d\sigma (x^{\prime}|u) \int d^4
y^{\prime}<[P_{\alpha \beta}
(x^{\prime},y^{\prime}),H_{in}(\tau ^{\prime})]>_{q \tau ^{\prime}} {%
\displaystyle \frac {\delta {\rho}_q( \tau ^{\prime}) }
{\delta <P_{\alpha \beta} (x^{\prime},y^{\prime})>_{\tau
^{\prime}} }} \ \}\mbox{ .}&
\end{eqnarray}
As shows our experience, the second addend in the right-hand side of Eq. (%
\ref{780}) removes usually all disconnected diagrams of a
given order from the set of diagrams which are supplied by the
first addend.

Some problems of this approach arise on the level of the
non-equilibrium thermodynamical Wick's theorem. However, there
is a set of methods which permits to overcome these
difficulties \cite{Z2, Fauser}.
c
\section{Conclusions}

In the present work, we have demonstrated an interesting
possibility for a covariant generalization of the real-time
Green's functions method which opens new ways after a
transition to the Wigner representation. In fact, this
generalization is based on the admissibility of introduction
in every point of the Minkowski space a unit time-like vector
constructed with the momentum vector of the corresponding
point of the phase space. Therefore, for a definition of the
preferred time-like direction in the Minkowski space, we used
a possibility contained in the Wigner function itself. The
relevant Assumptiones (Sec. 2) allow us to write, starting from
the Heisenberg representation, a motion equation in a
covariant form for the one-particle Wigner function (Sec. 2)
and, after that, to obtain a generalized RKE in the frame of
the Kadanoff-Baym technique (Sec. 3). Such a way is found
especially effective for a kinetic description of subsystems
with inner degrees of freedom and allows us to eliminate some
ambiguities of relativistic kinetic theory based on the
Kadanoff-Baym equations.

As a test of the suggested approach, the Fermi sector of the
well known Walecka model of the relativistic nuclear matter
was chosen (Sec. 3 and 4). For this model, RKE of the Vlasov
type (Sec. 3) and of the BUU type (Sec. 4) were obtained. In
the latter case, the collision integral was derived only for a
spin saturated system without antinucleon states. The
calculations were performed within the framework of the
quasiparticle approximation (QPA). We discuss also a
possibility of an extension of this approximation. An
agreement with a set of well-known simple results in the
literature denotes a noncontradictory character of the
proposed generalizations. It is safe to assume that methods
suggested here provide also a correct description of spin
degrees of freedom, states with negative energies (e.g., for
the annihilation channel in the theory of non-equilibrium
electron-positron plasma), and so on.

Finally, the covariant modification of the real-time Green's
functions method suggested in Sec. 3 was generalized to a
kinetic description of strongly non-equilibrium states (Sec.
5). For this aim, we suggest to use the non-equilibrium
statistical operator method.

The non-equilibrium relativistic nuclear matter represents a
good example of very complicated object for theoretical and
experimental investigations. Of course, the simplest toy model
and our present results discussed in Sec. 4 are only a rough
approximation of the reality. At present, a lot of actual
problems still remain in the relativistic kinetic theory,
e.g., going beyond the quasiparticle approximation, taking
into account nonlocal (and, in particular, non-Markovian)
effects in kinetic processes, cluster decomposition, etc.
(see, e.g., \cite{BM,CP,K,KM,R,BK,ARCH,MAL93,kremp}). An
important feature of the examined approach is a lack of some
dynamical constraints into a corresponding quantum field
theory (these constraints are introduced in order to eliminate
"unnecessary", non-physical degrees of freedom). The dynamical
introduction of these constraints into the Kadanoff-Baym
formalism is a rather non-trivial problem.
 A necessity to consider
such constraints creates serious problems for the kinetic
theory (e.g., in order to describe the $\Delta$-isobar
subsystem of nuclear matter). From our point of view, the
approach suggested in the present work can serve as a reliable
basis for such kind of further researches in the relativistic
kinetic theory.

\section*{ Acknowledgments}

We wish to thank V.~G.~Morozov for stimulating discussions and
S.~Mr\'owczy\'nski for addressing our attention to some
questions discussed here.

This work was partly supported by the State Committee of
Russian Federation for Higher Education under grant N 97-0-6.1-4
and was completed under the auspices of the U.S. Department of
Energy by the Los Alamos National
Laboratory under contract no. W-7405-ENG-36.

One from authors (S.A.S.) thanks the Soros Education Program
for support.

\section*{ Appendices}

\appendix

\section{ Proof of Eq. (25)}

Let us derive a proof of the Eq. (\ref{240}) using a boost transformation.

$$ \int d^3z^{\prime}\, S({\bf z} -{\bf
z}^{\prime},0)f(z^{\prime})\ =\ i\gamma ^0 f(z) \mbox{ ,}
\eqno (A1) $$ for an arbitrary function of the field operator
$f(z)$. The integration here is fulfilled over hyperplane of
the constant time, $z^0={z^{\prime}}^0$. Let us perform a
transition to an arbitrary frame of reference using a
transformation from the homogeneous Lorentz group $$
x^\mu\,=\,\Lambda^\mu_{\; \nu}(u)z^\nu , \quad
{x^{\prime}}^\mu\,=\Lambda^\mu_{\;\nu}(u){z^{\prime}}^{\nu}
\,.\eqno(A2) $$
The velocity vector, $u^\mu,$ defines simultaneously the
orientation of the
hyperplane, $\sigma(u)$. The transformation of the commutator function, $%
S({\bf z} -{\bf z}^{\prime},0),$ in Eq. (A1) is obtained using
the corresponding unitary operator, $U(\Lambda)$ \cite{sch},
$$ S({\bf z} -{\bf z}^{\prime},0)=S[\Lambda
(x-x^{\prime})]=U(\Lambda)S(x-x^{\prime})U^{-1}(\Lambda)\mbox{
,} \eqno(A3) $$
where $x,x^{\prime}\in \sigma(u)$. Similarly, we have: $$
f(z)=U(\Lambda)f(x)U^{-1}(\Lambda) \mbox{ .} \eqno(A4) $$
Let us take into account also the relation $$
U^{-1}(\Lambda)\gamma^\mu U(\Lambda)=\Lambda^\mu_{\; \nu }(u)\gamma^\nu %
\mbox{ .}\eqno(A5) $$
This equality is a consequence of the invariance under the
Lorentz transformation (A2) of the scalar product of $\psi(x)$
and $\phi(x)$ \cite {sch},
\[
(\psi ,\phi)_\sigma =\int d\sigma^\mu (x|u)\bar\psi(x)\gamma_\mu\phi(x)%
\mbox{ .}
\]
Finally, it is necessary to consider the equalities
$\Lambda^0_{\; 0}(u) =u^0 $ and $\Lambda^0_{\; k}(u)=u^k
(k=1,2,3)$. Formulas (A2)-(A3) allow us to realize the
transformation from Eq. (A1) to Eq. (\ref{240}). Another proof
of Eq. (\ref{240}) can be found in Ref. \cite{S2}.

\section{ Spectral properties of the modified Green's functions}

The spectral representations of the covariant contour Green's
functions introduced in Sec. 3 can be also considered in a
covariant form (we follow the work \cite{MH} were a similar
approach was fulfilled in the framework of a non-covariant
formalism). The main feature of these representations is the
selection of the time arguments in the form (\ref{150}). As an
example, let us consider the spectral decomposition of the
retarded and advanced Green's functions. According to the
definitions, (\ref{160}) and (\ref{360}), we have (see the
last remark in Sec. 3.2)
\[
G^{R/A}(xp)=\pm(2\pi)^{-4}\int dy
e^{ipy}\{G^>(x_+,x_-)-G^<(x_+,x_-)\}\theta [\pm y_\mu u^\mu]
\mbox{ ,}
\]
where $u^{\mu}$ is the unit vector (\ref{120}). From here, it
follows
\[
G^{R/A}(xp)=\pm\frac{1}{2\pi i}\int
d^4p^{\prime}\int\limits^\infty
_{-\infty}\! d\eta\, \delta^{(4)}(p^\mu-p^{\prime\mu}+u^\mu \eta)\frac{%
G^>(xp^{\prime})-G^<(xp^{\prime})} {\pm\eta
-i\varepsilon}\mbox{ ,}
\]
or $$
G^{R/A}(xp)=\pm \frac 12 [G^>(xp)-G^<(xp)]+\stackrel{\sim}{G}(xp)\mbox{ ,} %
\eqno(B1) $$ where $$ \stackrel{\sim}{G}(xp)=\frac{1}{2\pi
i}\int d^4p^{\prime}\int\limits^\infty
_{-\infty}\!d\eta\,\delta^{(4)}(p^\mu-p^{\prime\mu}+u^\mu
\eta)[G^>(xp^{\prime})-G^<(xp^{\prime})]\mathcal{P} \left(\frac{1}{\eta}%
\right) \mbox{ .} \eqno (B2) $$ Let us transform the last
relation using the decomposition of an arbitrary vector,
$a^\mu,$ on the time-line direction, $u_\mu$ (\ref{120}), and
orthogonal to it space-like component,
\[
a^\mu =u^\mu (au)+\Delta^\mu_\nu a^\nu \equiv a^\mu_\parallel + a^\mu_\perp %
\mbox{ ,}
\]
where the projection operator, $\Delta^{\mu\nu},$ is defined by Eq. (\ref{90}%
). Then, we can write the following representation $$
\delta^{(4)}(p^\mu-p^{\prime\mu}+u^\mu \eta)=\delta [n_\mu
(p^\mu-p^{\prime\mu}+u^\mu \eta)]\,\delta_\sigma
(p^\mu_\perp-p^{\prime\mu}_\perp )\mbox{ ,} \eqno (B3) $$ where
function $\delta_\sigma (p^\mu_\perp)$ is defined on the
space-like hyperplane $\sigma(u)$ . Using Eq. (B3), we can
rewrite Eq. (B2) in the final form: $$
\stackrel{\sim}{G}(xp)=\frac{1}{2\pi i}\int\limits^\infty_{-\infty}\!d\omega%
\,\mathcal{P}\left(\frac{1}{\omega -\sqrt{p^2}}\right)\{G^>(x;p_\perp,%
\omega) -G^<(x;p_\perp,\omega)\} \mbox{ .}\eqno (B4) $$ The
denominator, $(\omega -\sqrt{p^2}),$ in Eq. (B4) shows clearly
that function $\stackrel{\sim}{G}(xp)$ is just the
off-mass-shell part of the Green's functions, $G^{R/A}(xp)$.
As far as we have
\[
G^R(xp)+G^A(xp)=2 \stackrel{\sim}{G}(xp) \ ,
\]
and an analogous relation for the mass operator,
$\Sigma^{R/A}$, the last term in Eq. (\ref{490}) represents
indeed the off-mass-shell part of the collision integral
(\ref{480}).


\newpage

\begin{thebibliography}{99}

\bibitem{BM}  W. Botermans and R. Malfliet, \textit{Phys. Rep.} \textbf{198}
(1990) 115.

\bibitem{SC}  M. Sch\"{o}nhofen, M. Cubero, B. L. Friman, W. N\"{o}renberg,
and G. Wolf, \textit{Nucl. Phys.} \textbf{A572} (1994) 112.

\bibitem{MH}  
S. Mr\'{o}wczy\'{n}ski and U. Heinz, \textit{Ann. Phys.
(N.Y.)} \textbf{229} (1994) 1.

\bibitem{DP}  J. E. Davis and R. J. Perry, \textit{Phys. Rev. C} \textbf{43}
(1991) 1893.

\bibitem{B}  R. Bertoncini, \textit{Int. J. Mod. Phys. B} \textbf{6} (1992)
3441.

\bibitem{VB}  D. N. Voskresensky, D. Blaschke, G. R\"{o}pke, and H. Schulz,
\textit{Int. J. Mod. Phys. E} \textbf{4} (1995) 1.

\bibitem{CM}  W. Cassing and U. Mosel, \textit{Prog. Part. Nucl. Phys.}
\textbf{25} (1990) 235.

\bibitem{W}  Sh.-J. Wang, B.-A. Li, W. Bauer, and J. Randrup, \textit{Ann.
Phys. (N.Y.) }\textbf{209} (1991) 251.

\bibitem{S1}  S. V. Erokhin, A. V. Prozorkevich, S. A. Smolyansky, and V. D..
Toneev,\textit{\ J. Theor. Math. Phys.} \textbf{95} (1993) 74.

\bibitem{S2}  V. D. Toneev, A. V. Prozorkevich, and S. A. Smolyansky,
\textit{Heavy Ion Phys.} \textbf{3} (1996) 37.

\bibitem{CH}  E. Calzetta and L. Hu,\textit{\ Phys. Rev. D} \textbf{37}
(1988) 2778.

\bibitem{H}  P. A. Henning, \textit{Phys. Rep.} \textbf{253} (1995) 1.

\bibitem{RP}  J. Rau and B. M\"{u}ller, \textit{Phys. Rep.} \textbf{272}
(1996) 1.

\bibitem{SS}  P. J. Siemens, M. Soyer, G. D. White, L. J. Latto, and K. T.
R. Davies,\textit{\ Phys. Rev. C }\textbf{40} (1989) 2641.

\bibitem{DM}  A. Dellafiore and F. Matera, \textit{Phys. Rev. C} \textbf{44}
(1991) 2456.

\bibitem{S3}  S. G. Mashnik, A. V. Prozorkevich, S. A. Smolyansky, and G.
Maino, \textit{Nuovo Cim. A }\textbf{109} (1996) 1699;\\ G.
Maino, S. G. Mashnik, A. V. Prozorkevich, and S. A. Smolyansky
 \textit{Izv. Akad. Nauk, Ser. Fiz.} \textbf{62} (1998) 1004.

\bibitem{S5}  S. A. Smolyansky, A. V. Prozorkevich, S. Schmidt, D. Blaschke,
G. R\"{o}pke, and V. D. Toneev,  \textit{Int. J. Mod. Phys. E}
\textbf{7} (1998) 515.
\bibitem{S6}  A. V. Prozorkevich and S. A. Smolyansky, \textit{in} Selected
Topics of Nuclear Physics, p. 127, JINR, Dubna, Russia, 1995.

\bibitem{S7}  A. V. Prozorkevich, S. A. Smolyansky, and V. D. Toneev, JINR
preprint P 7-95-362, Dubna, Russia (1995).

\bibitem{S8}  A. V. Prozorkevich, S. A. Smolyansky, and V. D. Toneev,
\textit{Yad. Fiz. }\textbf{59} (1996) 804 [\textit{Phys. At. Nucl.} \textbf{%
59} (1996) 766].

\bibitem{S9}  G. Maino, S. G. Mashnik, S. A. Smolyansky, A. V. Tarakanov,
and D. Tocci, \textit{Izv. Akad. Nauk, Ser. Fiz.} \textbf{60} (1996) 58 [%
\textit{Bull. Russian Academy of Sci.} \textbf{60} (1996)
731].

\bibitem{S10}  S. A. Smolyansky, A. V. Prozorkevich, S. G. Mashnik, and G.
Maino,\textit{\ in} Proc. Int. Conf. on Nucl. Data for Science
and Technology, Trieste, Italy, May 19-24, 1997, SIF
Proceedings, vol.59, part I, p. 325, Bologna, 1997.

\bibitem{n}  P. T. Matthews, 
\textit{Phys. Rev. }\textbf{75} (1949) 1270; N. M.
Kroll,\textit{\ ibid}, 1321; V. G. Kadyshevsky, \textit{J.
Exp. Theor. Phys. (U.S.S.R.) }\textbf{46} (1964) 654
[\textit{Sov. Phys. JETP }\textbf{19} (1964) 443]; N. M.
Atakishev, R. M. Mir-Kassimov, and Sh. M. Nagiev,
\textit{Theor. Mat. Fiz.}
\textbf{32} (1977) 34; E. A. Dey, V. N. Kapshay, and N. B. Skachkov,\textit{%
\ ibid}, \textbf{101} (1994) 69.

\bibitem{Z2}  
D. N. Zubarev, V. G. Morozov, and G. R\"{o}pke, Statistical
Mechanics of Nonequilibrium Processes, Vol. 2, Akademie Verlag
GmbH, Berlin, 1997.

\bibitem{D}  P. Danielewicz, 
\textit{Ann. Phys. (N.Y.) }\textbf{152} (1984) 239.

\bibitem{G}  S. R. de Groot, W. A. van Leeuwen, and Ch. G. van Weert,
Relativistic Kinetic Theory, North-Holland, Amsterdam, 1980.

\bibitem{CP}  S. Chattopadhyay and D. Pal,\textit{\ J. Phys. G: Nucl. Part.
Phys.} \textbf{20} (1994) 357.

\bibitem{K}  H. S. K\"{o}hler,\textit{\ Phys. Rev. C} \textbf{51} (1995)
3232.

\bibitem{KB}  
L. P. Kadanoff and G. Baym, Quantum Statistical Mechanics,
Benjamin, New York, 1962.

\bibitem{SW}  B. D. Serot and J. D. Walecka, 
\textit{Adv. Nucl. Phys.} \textbf{16} (1986) 1.

\bibitem{ML}  J. A. McLennan, Non-Equilibrium Statistical Mechanics, New
Jersey, Prentice Hall, 1989.

\bibitem{Z1}  D. N. Zubarev, V. G. Morozov, and G. R\"{o}pke, 
Statistical Mechanics of Nonequilibrium Processes, Vol. 1,
Akademie Verlag GmbH, Berlin, 1996.

\bibitem{ZM}  J. Zimanyi and S. A. Moszkowski, \textit{Phys. Rev. C} \textbf{%
42} (1990) 1416.

\bibitem{BF}  P. Bernardos, V. N. Fomenko, N. V. Giai, M. L. Quelle, S.
Marcos, R. Niembro, and L. N. Savushkin, \textit{Phys. Rev. C}
\textbf{48} (1993) 2665; V. N. Fomenko, S. Marcos, and L. N.
Savushkin, \textit{J. Phys. G: Nucl. Part. Phys.} \textbf{19}
(1993) 545.

\bibitem{DC}  A. Delfino, C. T. Coelcho, and M. Malheiro, \textit{Phys. Rev..
C} \textbf{51} (1995) 2188.

\bibitem{KM}  V. M. Kolomiets, A. G. Magner, and V. A. Plujko, \textit{Yad.
Fiz.} \textbf{55} (1992) 2061 [\textit{Sov. J. Nucl. Phys.}
\textbf{55} (1992) 1143].

\bibitem{R}  G. R\"{o}pke, H. Schulz, K. K. Gudima, and V.D.Toneev, \textit{%
Fiz. Elem. Chastits At. Yadra} \textbf{21} (1990) 364
[\textit{Sov. J. Part. Nucl.} \textbf{21} (1990) 153].

\bibitem{BK}  Th. Bornath, D. Kremp, W. D. Kraeft, and M.Schlanges, \textit{%
Phys. Rev. E }\textbf{54} (1996) 3274.

\bibitem{sch}  S. S. Schweber, An Introduction to Relativistic Quantum Field
Theory, Row, Peterson and Co, New York, 1961.

\bibitem{LW}  N. P. Landsman and C. G. van Weert, \textit{Phys.Rep.} \textbf{%
145} (1987) 141.

\bibitem{BD}  J. D. Bjorken and S. D. Drell, Relativistic Quantum Fields, Mc
Graw-Hill, San Francisco, 1964.

\bibitem{Elze87}  H. -Th. Elze, M. Gyulassy, D. Vasak, U. Heinz, H. St\"{o}%
cker, and W. Greiner, \textit{Mod. Phys. Lett.} \textbf{A2}
(1987) 451.

\bibitem{ARCH}  
P. Kr\'{a}l, \textit{J. Stat. Phys.} \textbf{86} (1997) 1337.

\bibitem{MAL93}  H. S. K\"{o}hler and R. Malfliet, \textit{Phys. Rev. C}
\textbf{48} (1993) 1034.

\bibitem{Fauser}  
R. Fauser and H. H. Wolter,\textit{\ Nucl. Phys. A}
\textbf{600} (1996) 491.

\bibitem{kremp}  D. Kremp, M. Bonitz, W. D. Kraeft, and M. Schlanges,
\textit{Ann. Phys. (N.Y.)} \textbf{258} (1997) 320.

\bibitem{Z-book} D. N. Zubarev, Non-equlibrium Statistical
Thermodynamics, Plenum Press, New York, 1974.

\bibitem{MH92}
S. Mr\'{o}wczy\'{n}ski and U. Heinz, \textit{reprint
TPR-92-37}.

\bibitem{me}Yu. B. Ivanov, J. Knoll and D. N. Voskresensky,
 report
hep-ph/9807351;\\ P. Bozek, \textit{Phys. Rev. C} \textbf{56}
(1997) 1452.

\bibitem{he} B. M\" uller, report nucl-th/9807042;\\
K. Geiger and B. M\" uller, \textit{Heavy Ion Physics} \textbf{7} (1998) 207;\\
 K. Geiger, \textit{Phys. Rev. D} \textbf{54} (1996) 949;
\textbf{56} (1997) 2665.\\ 
D. Boyanovsky, H. de Vega, R.
Holman, S.Kumar, R. Pisarsky, report hep-ph/9802370.

\end{thebibliography}
\end{document}